\documentclass[pmlr,twocolumn,10pt]{jmlr}

\usepackage{indentfirst}
\usepackage{booktabs}
\usepackage{siunitx}
\usepackage[switch]{lineno}
\usepackage{url}
\usepackage{multicol}
\usepackage{graphicx}

\jmlrvolume{LEAVE UNSET}
\jmlryear{2024}
\jmlrsubmitted{LEAVE UNSET}
\jmlrpublished{LEAVE UNSET}
\jmlrworkshop{Machine Learning for Health (ML4H) 2024}

\title[NSTRI Global Collaborative Research Data Platform]{NSTRI Global Collaborative Research Data Platform}

\author{%
\Name{Hyeonhoon Lee}\Email{hhoon@snu.ac.kr}\\
\addr Seoul National University Hospital, Republic of Korea
\AND
\Name{Hanseul Kim}\Email{uo3359@snu.ac.kr}\\
\addr Seoul National University Hospital, Republic of Korea
\AND
\Name{Kyungmin Cho}\Email{kmcho1201@snuh.org}\\
\addr Seoul National University Hospital, Republic of Korea
\AND
\Name{Hyung-Chul Lee} \Email{vital@snu.ac.kr}\\
\addr Seoul National University Hospital, Republic of Korea
}

\begin{document}
\maketitle

\section{Introduction}
\label{sec:intro}

Access to valuable medical datasets from Korean hospitals has been significantly restricted for international researchers due to strict government regulations on health data sharing. The National Strategic Technology Research Institute (NSTRI) Data Platform addresses this challenge through Seoul National University Hospital's (SNUH) approved ICT regulatory sandbox status, enabling secure and compliant access to Korean healthcare pseudonymized data for global collaborative research. To overcome these limitations, our platform leverages SNUH's unique regulatory approval to enhance data accessibility and empower international researchers with advanced artificial intelligence (AI) tools, facilitating legitimate utilization of Korean healthcare data while maintaining compliance with data protection requirements.

The NSTRI Data Platform provides medical data from both SNUH and international healthcare institutions, enabling AI researchers to develop and validate machine learning models with enhanced equity and generalizability across diverse populations. This global collaborative research is powered by several key AI technologies: (1) intelligent data search engines that enable fast and accurate access to specific medical data, (2) automated medical translation tools that overcome language barriers, (3) drug search engines that facilitate cross-regional pharmaceutical research, and (4) an advanced large language models (LLM)-powered chatbot service that assists researchers with medical data exploration and analysis (\textbf{Appendix A} and \textbf{Figure 1}). These AI-empowered tools streamline tasks that were previously managed manually, significantly increasing research efficiency and data usability across different healthcare contexts. By providing researchers access to demographically diverse medical data through these automated tools, the platform ensures that developed AI models can be more reliable and applicable across different healthcare contexts and populations, addressing the critical need for reducing bias and improving generalizability in healthcare AI development. Through these capabilities, researchers and clinicians can efficiently locate and leverage the data they need, maximizing interaction across datasets with customisable search capabilities to meet specific user needs.

\section{Methods}
\label{sec:methods}

    \textbf{Data Pipeline:} The NSTRI Data Platform implements a comprehensive data pipeline that processes and integrates diverse medical data sources from SNUH's clinical data warehouse, including electronic medical records (EMR) with structured clinical data (patient demographics, ICD-10 diagnoses, medications, laboratory results, vital signs), medical imaging data (X-ray, CT, MRI with DICOM metadata), unstructured clinical texts (medical notes, radiology reports, pathology reports), and continuous monitoring data from medical devices. Our platform also provides international medical datasets from PhysioNet (including MIMIC-IV and eICU Collaborative Research Database) \cite{PhysioNet2019}. 

    \textbf{Dataset Search Engine:} 
    Our core search engine leverages semantic search technology powered by PubMedBERT-base-embeddings model \cite{PubMedBERT2020}, creating domain-specific embeddings that understand medical terms and expressions at an expert level than general-purpose models. This semantic understanding enables healthcare AI researchers to quickly find relevant data across our international medical datasets.
    
    \textbf{Medical Translator:}
    Built on the EEVE-Korean-v1.0 model \cite{EEVE2024}, this tool is specifically optimized for Korean-English medical translations. It employs a Transformer structure with memory optimization techniques like bitsandbytes for efficient float16 data loading. The model's sampling parameters (e.g., max tokens, temperature, top-p, top-k) are finely tuned to maintain accuracy and consistency, ensuring precise translation of complex medical terminology.
    
    \textbf{Drug Search Engine:}
    This specialized tool provides comprehensive data on drug names, main ingredients, and contraindications. It utilizes ATC codes to classify therapeutic purposes and relationships among similar drugs, supporting safer and more informed prescription decisions across different healthcare systems and regions.
    
    \textbf{LLM-powered Research Assistant:}
    Our advanced medical chatbot combines the Medical Terminology Mapper with retrieval augmented generation (RAG)-based technology for interactive guidance. It converts text into structured codes using standard medical coding systems (SNOMED-CT and LOINC) through a high-performance combination of PubMedBERT and RAG-based LLama model. The system uses ChromaDB for efficient data chunk retrieval and processes responses through the LLama-3.1-8B-Instruct model \cite{LLama2024}, delivering reliable, context-rich medical information based on user queries. This ensures a streamlined and precise terminology mapping experience for healthcare professionals while providing interactive assistance for data exploration.

    \textbf{Research Pod Architecture:} The platform operates within a secure cloud environment through a specialized research pod architecture. Each research pod implements multiple security layers: (1) SSL VPN-based access control that ensures authenticated entry only for authorized users, (2) Egress and Ingress controllers that prevent unauthorized data extraction while permitting access only to vetted machine learning libraries, and (3) a containerized environment that enables researchers to utilize security-assessed external libraries (pypi, conda-forge, cran) for analysis. This architecture maintains strict data protection while providing researchers the necessary computational resources and tools for their analysis. The research pods are deployed with pre-configured security policies that manage external API access and monitor data flow, ensuring compliance with institutional security requirements while facilitating efficient research workflows.

\section{Results}
\label{sec:results}

    \textbf{Datasets:} The NSTRI Data Platform provides access to 10 datasets from Seoul National University Hospital, each with specific access permissions, ensuring compliance with ethical and regulatory standards. The datasets are categorized by access type, and they can be joined using a primary key across datasets (\textbf{Appendix B}).

    \textbf{Research Projects:} The NSTRI Data Platform is actively supporting digital health data analysis through pilot deployments at SNUH and partner research institutions. Twenty-four project teams are currently utilizing the platform for medical machine learning research.

\section{Discussion}
\label{sec:discussion}
     
    \textbf{Challenges:} A primary challenge in developing the NSTRI Data Platform is harmonizing heterogeneous medical data across different institutions and countries. Each healthcare system uses distinct data formats, coding systems, and documentation practices, making standardization complex. For aligning multiple institutes, AI-empowered sophisticated preprocessing pipelines and extensive clinical validation is under development.

    \textbf{Lessons Learnt:} The importance of balancing security and accessibility when handling medical data became clear. This experience underscored the need to maintain researcher convenience while enhancing data security, providing insights into essential security and user-friendliness measures for future medical data platforms.

    \textbf{Future Plans for Improvement:} Moving forward, we plan to expand our dataset, improve search accuracy, and enhance collaborative capabilities with external organisations. By partnering with hospitals and research organisations, we aim to improve data quality, broaden application scope, and contribute to advancing digital healthcare by enabling wider research access.

\acks{This research was supported by a grant of the Korea Health Technology R\&D Project through the Korea Health Industry Development Institute (KHIDI), funded by the Ministry of Health \& Welfare, Republic of Korea (grant number: RS-2024-00403047)}

% Replace the bibliography section with this:

\appendix

\section{NSTRI Data Platform Architecture}\label{apd:first}

The NSTRI Data Platform is structured around a central \textit{Home} node, with major features organized into key sections:

\begin{itemize}
    \item \textbf{Datasets}: This section provides data search functionalities, allowing researchers to explore both datasets and relevant publications. It includes \textit{Datasets Search} and \textit{Papers Search}.
    \item \textbf{Tools}: A suite of tools designed to support medical data analysis. It includes:
        \begin{itemize}
            \item \textit{Drug Search} for drug-related information,
            \item \textit{Medical Translator} for Korean-English medical translations,
            \item \textit{Medical Terminology Mapper} for standardized medical term mapping, and
            \item \textit{LLM Chatbot \& API} for interactive data exploration and API access.
        \end{itemize}
    \item \textbf{Projects}: Supporting research collaboration through features like the \textit{Research Pod}, which provides a dedicated space for research activities, and \textit{Projects}, which enables project management and organization.
    \item \textbf{Codes}: Through \textit{GitLab}, offering \textit{Code Space} and a \textit{Community} hub that empower researchers to manage code, share their projects, and participate in discussions.
    \item \textbf{Supports}: Providing user support, including a comprehensive \textit{User Manual} to assist with platform usage.
\end{itemize}

This structured architecture is designed to enhance data accessibility and support research activities effectively.

\section{NSTRI Datasets Details}\label{apd:second}

\begin{itemize}
        \item \textbf{Credentialed Access}: Providing summary information only. Full access requires individual IRB/DRB approval and is limited to approved Research Pod.
            \begin{itemize}
                \item \textbf{SNUH CDM}: OMOP-CDM dataset derived from SNUH’s EMR data.
                \item \textbf{SNUH NOTE}: Pseudonymized clinical notes from SNUH.
                \item \textbf{VitalDB}: Pseudonymized biosignal data from 200,000 surgical patients.
                \item \textbf{SNUH MACCE}: Dataset of 288,172 preoperative ECG XML files with MACCE labeling.
                \item \textbf{SNUH CXR}: Chest X-ray dataset from SNUH.
                \item \textbf{INSPIRE 300K}: Pseudonymized dataset of 300,000 surgical patients.
            \end{itemize}
        \item \textbf{Restricted Access}: Datasets restricted to platform use only. No external export permitted.
            \begin{itemize}
                \item \textbf{LYDUS ECG 160K}: Curated dataset of 167,199 12-lead ECG exams with machine and human expert labeling.
            \end{itemize}
        \item \textbf{Open Access}: Freely downloadable datasets for external use.
            \begin{itemize}
                \item \textbf{INSPIRE 150K}: Publicly accessible subset (50\%) of the INSPIRE dataset.
                \item \textbf{ICU ARREST}: ICU lead-II ECG \& PPG (48-hour) dataset with normal and cardiac arrest labeling (6,102 ICU stays).
                \item \textbf{LYDUS ECG 50K}: Curated dataset of 50,000 12-lead ECG exams with machine and human expert labeling.
            \end{itemize}
    \end{itemize}

\onecolumn
\begin{figure*}[!hb]
    \centering
    \includegraphics[width=1.0\textwidth]{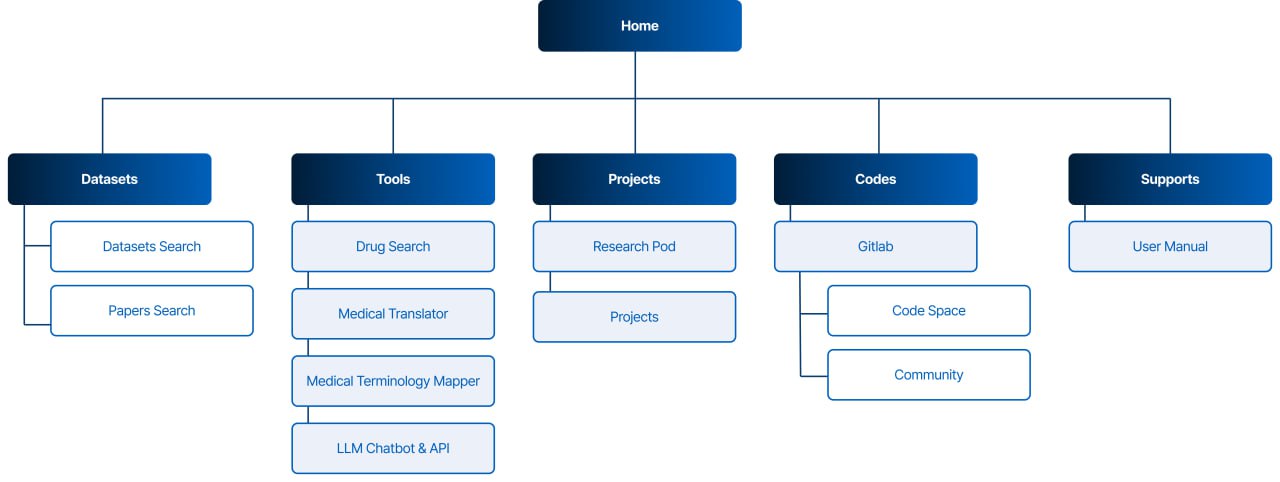}
    \caption{NSTRI Data Platform Architecture}
    \label{fig:enter-label}
\end{figure*}


\begin{thebibliography}{9}

\bibitem[Moody et al.(2001)]{PhysioNet2019}
Moody, G.B., Mark, R.G., \& Goldberger, A.L. (2001).
\newblock {PhysioNet: a Web-based resource for the study of physiologic signals}.
\newblock {\em IEEE Engineering in Medicine and Biology Magazine},
  20(3):70--75.

\bibitem[Gu et al.(2021)]{PubMedBERT2020}
Gu, Y., Tinn, R., Cheng, H., Lucas, M., Usuyama, N., Liu, X., Naumann, T., Gao, J., \& Poon, H. (2021).
\newblock {Domain-Specific Language Model Pretraining for Biomedical Natural Language Processing}.
\newblock {\em ACM Transactions on Computing for Healthcare},
  3(1):1--23.

\bibitem[Kim et al.(2024)]{EEVE2024}
Kim, S., Choi, S., \& Jeong, M. (2024).
\newblock {Efficient and effective vocabulary expansion towards multilingual large language models}.
\newblock {\em arXiv preprint arXiv:2402.14714}.

\bibitem[Dubey et al.(2024)]{LLama2024}
Dubey, A., Jauhri, A., Pandey, A., et al. (2024).
\newblock {The llama 3 herd of models}.
\newblock {\em arXiv preprint arXiv:2407.21783}.

\end{thebibliography}
\end{document}